\documentstyle[preprint,aps]{revtex}
\draft
\begin{document}
\title{Optical absorption and single--particle excitations
in the 2D  Holstein t--J model}
\author{B. B\"auml, G. Wellein and H. Fehske } 
\address{Physikalisches Institut, Universit\"at Bayreuth,
D--95440 Bayreuth, Germany}
\date{Bayreuth, 17 February 1998}
\maketitle
\input{epsf}
\def\gsim{\hbox{$\lower1pt\hbox{$>$}\above-1pt\raise1pt\hbox{$\sim$}$}}
\def\lsim{\hbox{$\lower1pt\hbox{$<$}\above-1pt\raise1pt\hbox{$\sim$}$}}
\def\cH{{\cal{H}}}
\def\ep{\varepsilon_p}
\def\om{\omega_0}
\begin{abstract}
To discuss  the interplay of electronic and lattice degrees of freedom in 
systems with strong Coulomb correlations we have performed an  
extensive numerical study of the two--dimensional Holstein t--J model.
The model describes the interaction of holes, doped in a quantum 
antiferromagnet, with a dispersionsless optical phonon mode.
We apply finite--lattice Lanczos diagonalization, combined 
with a well--controlled phonon Hilbert space truncation, to the
Hamiltonian. The focus is on the dynamical properties.
In particular we have evaluated the single--particle spectral 
function and the optical conductivity for characteristic 
hole--phonon couplings, spin exchange interactions 
and phonon frequencies. The results are used to analyze the 
formation of lattice hole polarons in great detail. 
Links with experiments on layered perovskites are made.  
Supplementary we compare the Chebyshev recursion and maximum 
entropy algorithms, used for calculating spectral functions, 
with standard Lanczos methods.
\end{abstract}
\thispagestyle{empty}
\newpage
\section{Introduction}
There has been renewed interest in the analysis of strongly coupled 
electron--phonon (EP) systems since it was recognized that intrinsic 
polaron--like lattice distortions are a common feature of 
several important classes of perovskites such as 
the high--$T_c$ cuprates~\cite{BEMB92,SAL95}, the non--metallic 
nickelates~\cite{CCC93}, and the colossal magnetoresistance 
manganites~\cite{MLS95,ELM97}. In these compounds,
e.g., in $\rm La_{2-x}Sr_x[Cu/Ni]O_4$ or $\rm La_{1-x}Ca_xMnO_3$, 
the charge carriers susceptible to lattice polaron formation 
mediate the antiferromagnetic (AFM) or ferromagnetic (FM) interactions
between the $\rm Cu^{2+},\,Ni^{2+}$ ($S=1/2,\,1$) or $\rm Mn^{3+}$ 
($S=3/2$) ions (core spins), suggesting that the local 
lattice or Jahn--Teller distortions play also a significant  
role in determining the electronic and magnetic 
properties~\cite{Trea95,ZCKM96}. 

A number of experimental observations indicate spatial modulations
of the spin and charge density in the copper--, nickel-- and 
manganese--based perovskites. For example,
dynamical stripe correlations of spins and holes were found from neutron 
scattering~\cite{Tr97} and X-ray absorption (EXFAS)~\cite{Biea96}
in superconducting $\rm La_{1.85}Sr_{0.15}CuO_4$ and $\rm YBa_2Cu_3O_{6.6}$. 
Rather static polaronic superlattice structures 
have been detected by electron~\cite{Chea94}, neutron~\cite{TBSL94} and
X--ray diffraction~\cite{Ziea97} measurements in hole doped (insulating)  
$\rm La_{2-x}Sr_xNiO_4$. Local structure studies of $\rm La_{1-x}Sr_xMnO_3$,
performed using pulsed neutron powder diffraction and 
atomic pair--density function analysis~\cite{Loea97}, suggest  
that nonuniform charge distributions related to lattice polarons 
exist in the paramagnetic, AFM and FM phases, and are even present in 
the metallic state up to $x\sim 0.35$ . 

A striking feature of all these charge and spin ordering phenomena 
is their cooperative nature, demonstrating the strong interaction between 
spin, charge and lattice degrees of freedom~\cite{Trea95,Area96}.  
Accordingly, the lattice anomalies are well reflected in the transport 
properties. For example, there is observed a striking suppression of 
superconductivity in the $\rm [La,Sr]_2CuO_4$ family compounds 
with $\sim 1/8$ doping, which is believed to have a 
connection with charge and spin stripe formation~\cite{Trea95}.  
Undoubtedly, structural correlations and  (polaronic) charge order are 
shown to be the driving force behind the pronounced resistivity
anomalies found in $\rm [La,Sr]_2NiO_4$ for hole concentrations 
1/3 and 1/2~\cite{Chea94,LC97}. 

Since polaronic effects might be responsible for some of the puzzling 
thermodynamic and transport properties of the cuprates, nickelates and 
manganites, increasing effort has been aimed to test the polaronic nature  
of charge carriers by means of optical experiments. One of the physical
quantities which contains valuable information about the low--energy  
excitations in interacting EP systems is the optical conductivity, 
$\sigma(\omega)$, usually determined from reflectivity measurements.
Here the anomalous broad optical absorption detected 
for the $\rm La_{2-x}Sr_x[Cu,Ni]O_4$ compounds 
in several (normal--state) infrared experiments has attracted a lot 
of attention. In particular for the nickelate system, where the existence
of self--trapped charge carriers is well established, the 
mid--infrared absorption band has been interpreted successfully in terms
of photon--assisted hopping of small--size lattice polarons~\cite{BE93,Caea96}.
On the other hand, from a comparison of the optical absorption of several 
perovskites ($\rm La_{1.67}Sr_{0.33}NiO_4$, 
$\rm Sr_{1.5}La_{0.5}MnO_4$, and $\rm Nd_{1.96}Ce_{0.04}CuO_4$),
it has been argued that in the low--doping cuprates the lattice
polarons appear to be rather large~\cite{Caea97,Emi93}. 
Although recent time--resolved optical relaxation measurements on 
$\rm YBaCuO_{7-\delta}$ give additional evidence for the 
existence of polaronic carriers in the cuprates~\cite{Miea97}, 
it is worth mentioning that the present experimental situation 
certainly does not permit to draw any decisive conclusion
concerning the relevance of the polaron/bipolaron scenario
for high--$T_c$ superconductivity. 
 
From a theoretical point of view, the main problem for a 
proper treatment of such systems, exhibiting besides a substantial
EP coupling strong Coulomb correlations, is that the coherent
motion of charge carriers, heavily dressed, e.g., by interaction 
with magnons, takes place on a strongly reduced energy
scale being almost comparable to the relevant phonon frequencies. 
As a result the standard strong--coupling Migdal--Eliashberg approach 
based on the adiabatic Migdal theorem might break down~\cite{RHF92,KMKF96},
and it has been argued that non--adiabatic vertex corrections are responsible
for the enhancement of $T_c$~\cite{PSG95,GPS95,GCP97}. Furthermore,
as a consequence of the pre--existing  magnetic ``self--localization'' 
of the charge carriers, polaronic effects play an important role 
even at rather moderate EP coupling 
strengths~\cite{FRWM95,WRF96,IET97,SPS97,DKR90}.
Therefore the investigation of such highly correlated EP systems
is of fundamental importance, not only in connection with the high--$T_c$
problem. 

Our aim here is to study what is perhaps the minimal microscopic
model with respect to a strong spin--charge--lattice coupling,
the so--called Holstein t--J model (HtJM). The focus is  
on the spectral properties of the HtJM Hamiltonian in the  
single--hole subspace relevant to the physically most interesting 
low doping regime. In order to obtain reliable results
for the complete range of exchange interactions, EP couplings 
and phonon frequencies we will employ numerical exact diagonalization
(ED) techniques on finite lattices, using massive parallel computers
such as the CRAY T3E. 

The paper is organized as follows: In the next section we  
introduce the HtJM and comment on previous numerical investigations.
Sec.~III presents our ED results for the  single--particle spectral 
function, which will be used in Sec.~IV to discuss the hole--polaron 
band formation. In  Sec.~V we analyze the optical response in the HtJM.
Our main results are summarized in Sec.~VI. A comparative study of several 
computational techniques proposed for calculating spectral 
properties of very large Hamiltonian matrices is performed in the Appendix.
\section{The Holstein \mbox{t}--J model}
Our model is the standard two--dimensional t--J model 
appended by an additional Holstein--like interaction 
term with a dispersionsless (optical)
phonon branch:
\begin{eqnarray}
{\cal H}=&&-t \sum_{\langle i j \rangle \sigma} 
\Big(\tilde{c}_{i\sigma}^\dagger 
\tilde{c}_{j\sigma}^{} + {\rm H.c.}\Big)+ J \sum_{\langle i j\rangle}
\Big(\vec{S}_i^{}\vec{S}_j^{} - \frac{1}{4}\tilde{n}_i^{}\tilde{n}_j^{}\Big)
\nonumber\\
&&\qquad- \sqrt{\ep\hbar\om}  \sum_i \big(b_i^{\dagger} + b_i^{} \big)\,
\tilde{h}_i^{}
\;+\;\hbar\om \sum_i \big(b_i^{\dagger} b_i^{} + \frac{1}{2}\big)\,.
\label{htjm}
\end{eqnarray}
The Holstein t--J Hamiltonian~(\ref{htjm}) acts in a projected Hilbert space
without doubly occupied sites, where $\tilde{c}_{i\sigma}^{(\dagger)}=
c_{i\sigma}^{(\dagger)}(1-\tilde{n}_{i,-\sigma}^{})$ is a 
spin--$\sigma$ electron annihilation (creation) operator on Wannier
site $i$, $\tilde{n}_{i}=\sum_{\sigma}
\tilde{c}_{i\sigma}^{\dagger}\tilde{c}_{i\sigma}^{}$, and 
$\vec{S}_i=\sum_{\sigma,\sigma^{\prime}}\tilde{c}_{i\sigma}^{\dagger}
\vec{\tau}^{}_{\sigma\sigma^{\prime}}\tilde{c}^{}_{i\sigma^{\prime}}$.
Within an effective single--band description,    
the parameters $t$ and $J$ measure the  transfer amplitude and the 
antiferromagnetic exchange interaction between  
nearest--neighbour sites on a square lattice, where
$t>J$ corresponds to the situation in the cuprates. For example,  
$J/t\simeq 0.4$ with $t\simeq 0.3$~eV is commonly used to model
the $\rm La_{2-x}Sr_xCuO_4$ system. In~(\ref{htjm}), we have chosen
the coupling to the hole, $\tilde{h}_i=1-\tilde{n}_i$ is the 
local density operator of the spinless hole, as the dominant
source of the EP coupling, because in the t--J model the unoccupied 
site corresponds to the Zhang--Rice singlet (formed 
from $\rm Cu \,3d_{x^2-y^2}$ and $\rm O\, 2p_{x,y}$ hole orbits)
for which the coupling should be much stronger than for the occupied
($\rm Cu^{2+}$) site~\cite{FRMB93,Fer94,SS95}.  
The hole--phonon coupling constant is denoted by $\ep$,  
and $\hbar\om$ is the bare phonon frequency (below $\hbar=1$ 
and all energies are measured in units of $t$). 
Then $q_i= (b_i^{\dagger}+b_i^{})/\sqrt{2M\om}$ 
may  be thought of representing an internal vibrational 
degree of freedom of the lattice site $i$ ($b_i^{(\dagger)}$ 
annihilates (creates) an optical phonon). Indeed for $\rm La_2[Cu,Ni]O_{4}$ 
the oxygen vibrational modes have a small dispersion in the Brillouin zone, 
i.e., they are revealed to be very localized in real space and 
can be considered as independent~\cite{PSSH96}. 
  
Whereas the electronic properties of the pure t--J model have been
studied extensively in the context of high--$T_c$ 
superconductivity~\cite{Da94}, very little is known theoretically 
about the interplay between magnetic exchange interaction and EP coupling
in the framework of the 2D HtJM. The changes of the quasiparticle (QP) 
properties due to the cooperative/competitive effects of the 
hole--phonon and hole--magnon interactions are expected to be 
very complex and as yet there exist no well--controlled 
analytical techniques to solve the problem particularly
in the physically relevant coupling regime, where the characteristic
electronic ($t,\;J$) and phononic ($\ep,\;\om$) energy scales are not well
separated. Naturally such a dressed hole QP will show the 
characteristics of both ``lattice'' and ``magnetic'' (hole) 
polarons~\cite{poldef}.   
 
Early attempts to study lattice polaron effects in the HtJM were based
on a variational squeezed--polaron approach to the phonon 
subsystem~\cite{DKR90}. The effective electronic Hamiltonian, 
obtained in the transformed phonon vacuum state, was 
treated by means of Gutzwiller or slave boson techniques and the main features 
of the ground state phase diagram of the Holstein Hubbard/t--J models
were worked out on a mean--field level of approximation~\cite{TFDB94,Feea94}.
Analyzing the static charge susceptibility of the HtJM, a transition to 
a Peierls distorted phase was found to occur at quarter filling~\cite{FD94}.
In the low--doping regime and for moderate EP interactions 
the lowest polaron--magnon band states were determined for the 
Lang--Firsov transformed 2D HtJM within 
the zero--phonon and spin--wave approximations, 
indicating the formation of large--size (mainly magnetic) hole polarons with 
soliton--like--shaped highly asymmetric wave functions, 
oriented along the diagonals of the square lattice~\cite{SSS91}.   
Combining the spin--wave approximation with an iterative Lanczos algorithm, 
the ground--state properties of the 2D HtJM were examined including 
multi--phonon states~\cite{SSS94}. A sharp crossover to small--size 
less--dispersive symmetric lattice hole polarons was obtained at a critical  
value of the EP coupling~\cite{SSS94}. Using the self--consistent 
Born approximation, the phonon--induced mass renormalization of a 
hole in the HtJM was found to depend strongly on the 
confining string potential due to the spin background~\cite{RHF92}.
Whereas a large AFM exchange interaction certainly enhances 
the tendency towards lattice polaron formation, there is evidence that 
in the limit $J\to 0$  the influence of phonons becomes even smaller than
in the corresponding uncorrelated (spinless fermion) model~\cite{RHF92}.  
More recently, the single--hole spectral density function and 
the optical conductivity were studied within the non--crossing 
approximation for spin--wave and phonon interactions
in a Holstein t--J--like Hamiltonian~\cite{KMKF96}. 
Perhaps surprisingly, the numerical estimate of the lowest order
vertex corrections to the hole--phonon coupling vertex 
gives relatively small values, indicating that Migdal's approximation
can be used in calculating the hole self--energy. 
This effect has been attributed to the significant suppression of
the QP residue for a correlated electron system~\cite{KMKF96}.
 
Of course these analytical approaches provide some insight into the problem,
but it is difficult to judge how close to the actual properties of the
HtJM the results are. From this point of view it is of interest to carry out 
exact diagonalizations of the HtJM on small clusters. Along this line, 
a number of ED studies have been performed treating the phonons in the
static limit, i.e., classically, in the so--called adiabatic or 
frozen phonon approximation~\cite{ZS92,FRMB93,RFB93,RFS94}. 
The stability of the ground state against static lattice distortions
was investigated and the phase diagram of the 2D HtJM has been explored 
in the one-- and two--hole sectors~\cite{FRMB93}, as well as for the 
quarter--filled band case~\cite{RFS94}, where a polaronic superlattice 
was found to occur above a critical EP coupling strength.    
Neglecting, however, the phonon dynamics, certainly overestimates the tendency
towards lattice polaron or charge--density--wave 
(CDW) formation~\cite{Fer96,WFBB97}.
As a first step to overcome the static treatment of the HtJM, a variational
Lanczos diagonalization technique has been proposed~\cite{FRWM95}, 
which is based on an inhomogeneous modified Lang--Firsov transformation
and allows for the description of static displacement field, non--adiabatic
lattice polaron and squeezing effects on fairly large clusters.
Extending the ED approach to the full quantum phonon model requires
a truncation of the bosonic part of the Hilbert space, which, 
strictly speaking, is infinite dimensional. 
Including at each site besides the phonon vacuum a 
second (coherent) displaced oscillator state, the 1D HtJM has been analyzed 
with respect to the occurrence of CDW order,  
superconducting fluctuations, and phase separation~\cite{Fer96}.  
Whether the complex (phonon--assisted) transport in the HtJM 
can be described with sufficient accuracy within such a simple 
two--state variational approach remains questionable, at least in
the intermediate coupling regime. For the 2D case 
more advanced truncation procedures in momentum space, 
retaining a single or small number of phonon 
modes~\cite{PSSH96,DGKR95}, have been employed to show that dynamical
in--plane phonons could help to stabilize inhomogeneous stripe--like
phases.  Taking into account all dynamical phonon modes, 
in a previous work~\cite{WRF96} the authors have examined
the ground--state properties of the 2D Holstein t--J model 
in the one-- and two--hole subspaces by means of direct Lanczos 
diagonalization. Most notably, the transition from delocalized 
to self--trapped lattice hole polarons (hole bipolarons), 
signaled by a suppression of the kinetic energy 
and an enhancement of the hole--phonon (hole--hole) correlation functions,  
was found to be favored by strong Coulomb correlations. 

In the present paper, we extend this rigorous approach, which preserves 
the full dynamics and quantum nature of phonons, to the calculation 
of dynamical quantities.  
In particular we compute the single--hole spectral function   
and the optical conductivity to discuss the  transport of hole polarons 
in the framework of the 2D HtJM. The technical details of the computational
procedures such as the basis implementation and symmetrization, 
the phonon Hilbert space truncation and the spectral moment approach 
will be discussed in the Appendix.  
\section{Single--particle spectral properties}
In the numerical analysis of the 2D HtJM we start with a discussion of the 
single--hole spectral function
\begin{equation}
A_{\vec{K}}(\omega)=\sum_{n,\sigma} 
\left|\langle {\mit \Psi}_{n,\vec{K}}^{(N-1)}
|\tilde{c}^{}_{\vec{Q}-\vec{K},\sigma}| {\mit \Psi}_{0,\vec{Q}}^{(N)}\rangle
\right|^2\;
\delta\left[\omega-(E_{n,\vec{K}}^{(N-1)}
-E_{0,\vec{Q}}^{(N)})\right]\,.
\label{akw}
\end{equation}
Figure~1 displays $A_{\vec{K}}(\omega)$ for the allowed values of the
momentum $\vec{K}$ of a ten--site square lattice. 
To visualize the intensities (spectral weights) connected with the various 
peaks (excitations) in each $\vec{K}$--sector we also have shown 
the integrated density of states
\begin{equation}
N(\omega)=\int_{-\infty}^{\omega}d\omega^{\prime}\frac{1}{N}
\sum_{\vec{K}}A_{\vec{K}}(\omega^{\prime})\,.
\label{nw}
\end{equation}  
Of course, in the absence of EP coupling we reproduce 
the single--particle spectrum of the pure t--J model, which  
for comparison is shown in Fig.~1~(a). 
In this case the single--hole excitations 
are reasonably well understood~\cite{Da94}.
$A_{\vec{K}}(\omega)$ contains a ``quasiparticle''  
peak at the bottom of the spectrum followed by a lump
of spectral weight with some internal structure. The QP pole, 
corresponding directly to the coherent single--hole ground state with
momentum $(3\pi/5,\pi/5)$,  
is separated by a pseudogap of approximative size $J$ from the lower edge
of a broad incoherent continuum $6t$ wide. Despite the significant
redistribution of intensity to higher energies as a 
consequence of multiple spin--wave (spinon) scattering of 
the magnetic polaron~\cite{Br95}, the QP residue still contains a substantial 
amount of spectral weight. It has been argued, that the 
QP peak and the low--energy satellite structures, which
can be identified as the next excited states of the string problem,
will survive in the bulk limit~\cite{Da94}. The momentum dependences of 
$A_{\vec{K}}(\omega)$ indicate  that the low--energy structure
of the $\vec{K}$--integrated spectrum is dominated by excitations 
with wave vectors near the magnetic Brillouin zone boundary (particularly 
by the $\vec{K}$--vectors closest to $(\pi/2,\pi/2)$ and 
$(\pi,0)$)~\cite{PZSD93}. 
On the other hand, for momenta $\vec{K}=(0,0)$ and  $(\pi,\pi)$ 
most of the weight of $A_{\vec{K}}$ is concentrated 
at high energies, i.e., in the incoherent part of the spectrum.

Now we consider the influence of an additional weak EP interaction 
(Fig.~1~(b)). In the weak--coupling regime 
the mass renormalization of the coherent QP band due to the
hole--phonon coupling is generally small compared with 
that arising from the hole--spin interactions (magnetic polaron regime). 
In particular,  the integrated density of states 
are barely changed from those of the pure t--J model. 
The new structures, nevertheless observed in the $A_{\vec{K}}$ spectra
shown Fig.~1~(b), correspond to predominantly ``phononic'' side 
bands separated from the particle--spin excitations by multiples of the  
bare phonon frequency $\omega_0$. However, these phonon resonances 
have less and less ``electronic'' spectral weight the more
phonons are involved. This is because $A_{\vec{K}}(\omega)$ measures 
the overlap of these excited states with the state obtained 
by creating a hole in the {\it zero--phonon} Heisenberg 
ground state.     

To analyze the formation of lattice hole polarons in the
2D HtJM, we have compared the non--adiabatic weak-- and 
strong--coupling situations in Fig.~2. Looking first
at the case $J=0.4$, panels~(a) and~(c), we observe that with
increasing $\ep$ the lowest peaks in each $A_{\vec{K}}$
start to separate from the rest of the spectrum. These states become
very close in energy and finally a narrow well--separated lattice 
hole--polaron band evolves in the strong--coupling case ($\ep=4.0$). Since 
the gap to the next energy ``band'' is of the order of     
$\omega_0$ (see panel~(c)), these excitations will be triggered by
an one--phonon absorption process. The transition to the 
lattice hole--polaron state is accompanied by a strong increase in 
the on--site hole--phonon correlations~\cite{WRF96},
indicating that the lattice polaron QP comprising 
a ``quasi--localized'' hole and the phonon cloud is mainly
confined to a single lattice site (small--size lattice hole polaron).
Compared to the non--interacting single--electron   
(Holstein model)~\cite{FLW97} or spinless fermion 
(Holstein--t model)~\cite{FRWM95} problems, 
the critical EP coupling strength for lattice polaron formation is 
considerably reduced due to magnetic prelocalization effects.
The phonon distribution function shows that 
the small--size lattice hole--polaron state is basically a multi--phonon state
(see Ref.~\cite{FWBB97}; cf. also Fig.~8 in the Appendix). 
For example, at $\ep=4.0$ and $J=0.4$ we found the mean phonon number 
in the ground state to be about 4--5. Here the phonons
will heavily dress the hole and the QP pole strength
becomes strongly suppressed (cf. the discussion in Sec.~IV). 
At the same time spectral weight is transferred to the 
high--energy part and the whole spectrum
becomes incoherently broadened. Therefore, we observe an 
overall smoothing of $N(\omega)$. We note that the vertex corrections 
to the hole--phonon vertex are non--negligible for the large 
hole--phonon couplings and intermediate phonon frequencies considered here  
($\ep$, $\sqrt{\ep\omega_0}\gg J$). This has to be contrasted with the weak
to moderate EP coupling regimes, where the vertex correction was found to be 
much smaller than unity~\cite{KMKF96}.

Next, we want to discuss the effect of the exchange interaction $J$.
For the t--J model it is well known that as long as $J/t$ is large, i.e.,
the spins are antiferromagnetically ordered, 
the delocalization of a hole is hindered as a consequence of a 
linear confining string potential originated from the 
misaligned spins along the hole hopping paths. 
If $J$ tends to zero, the string potential is weakened and
the hole performs large excursions polarizing the spin 
background ferromagnetically. In a finite system 
this so--called Nagaoka transition takes place at a {\it finite} 
critical exchange interaction ($J_N=0.13807$ for the ten--site lattice).
For $J<J_N$, the single--hole ground state has momentum $\vec{K}=(0,0)$ 
and is fully spin aligned. At $J=0$, the spectrum is completely 
incoherent, extending over almost the unperturbed bandwidth $8t$. 
As can be seen from Fig.~2~(b), showing the case $J=0.1<J_N$, 
a weak EP coupling does not affect this scenario. 
Quite the reverse, from a self--consistent Born calculation
it was deduced that the influence of phonons on such a
``Nagaoka spin--polaron'' becomes even smaller than in the free fermion 
case~\cite{RHF92}. The situation, however, is changed completely 
in the strong EP coupling regime. As $\ep$ increases, the mobility of the hole
is reduced and finally a strong local lattice distortion traps the hole
almost on a single site. The hole--trapping makes even a small AFM exchange 
interaction very effective and we observe, as for the static ($\omega_0\to 0$)
HtJM~\cite{RFB93}, an enhancement of the AFM correlations in 
the spin background. Consequently, the whole spectrum 
looks very similar to that obtained for $J=0.4$ (cf. panels (c) and~(d)).  
The only difference is that now the first spin excitations,
having an excitation energy $\propto J^{(2/3)}$~\cite{Daea90},
occur just above the lattice hole--polaron band, thus filling the gap
between the first and second polaronic bands.

Finally, we consider the regime, where the bare phonon frequency
is much larger than the effective hole bandwidth 
of the t--J model. For $0.1\leq J \leq 0.4$, the coherent bandwidth 
can be approximated by  ${\mit \Delta}E^{t-J}\simeq 1.5 J^{\alpha}$ 
with $\alpha\simeq 1$~\cite{EBS90}.     
Figure~3 shows the results for the spectral function 
$A_{\vec{K}}(\omega)$ at $\omega_0=3.0$, $\ep=4.0$, and $J=0.4$.
In this case  the lowest (QP) excitation band is only weakly 
renormalized by the phonons which can follow the motion 
of the magnetic polaron instantaneously. The weights of the 
first $m$--phonon states in the ground state 
are 0.436, 0.332, 0.15, and 0.055 for $m=0$, 1, 2, and 3, respectively. 
As can be seen by comparing Figs.~1~(a), 2~(c) and~3, the QP pole 
strength is determined by the subtle interplay of $\ep$ and $\omega_0$. 
In the anti--adiabatic strong--coupling limit, the high--energy part of 
the $A_{\vec{K}}(\omega)$ spectrum is clearly dominated 
by interband transitions according to multi--phonon 
absorptions (cf. the pronounced jumps in $N(\omega)$ 
at energies $\omega\simeq n\times\omega_0$).
\section{Polaron band formation}
To illustrate the formation of the lattice hole--polaron 
band in some more detail,
we have calculated the ``coherent'' band dispersion of the 2D HtJM, 
$E_{\vec{K}}$, on a 16--site lattice. $E_{\vec{K}}$ was determined 
from the lowest pole of each $A_{\vec{K}}(\omega)$ having finite 
spectral weight. The resulting band structure is shown in Fig.~4 
along the principal directions in the Brillouin zone. 
For the pure t--J model, the dispersion of the QP band  
is highly consistent with a simple tight--binding form~\cite{MH91} that
includes hole hopping processes to first and second--nearest neighbours 
on the same sublattice only (leaving the magnetic 
order undistorted). From Fig.~4~(a) the minima of the QP dispersion
are found to be located at the momenta $\vec{K}=(\pm\pi/2,\pm\pi/2)$ 
(the hidden symmetry of the 4$\times$4 cluster leads to an accidental
degeneracy with the  $\vec{K}=(\pm\pi,0)$, $(0,\pm \pi)$ states).  
Obviously, the energy dispersion is not significantly 
changed  including a weak EP coupling,
provided that the phonon frequency exceeds 
the effective bandwidth of the magnetic polaron 
($\omega_0=0.8\geq {\mit \Delta}E^{t-J}$). On the other hand,
in the adiabatic regime ($\omega_0\ll {\mit \Delta}E^{t-J}$), 
a remarkable ``flattening'' of the band structure of the HtJM takes place, 
which can be attributed to the ``hybridization'' of the 
hole QP band with the dispersionsless optical phonon. As a result the 
coherent bandwidth ${\mit \Delta}E=\sup_{\vec{K}}E_{\vec{K}}
-\inf_{\vec{K}}E_{\vec{K}}$ is roughly given by $\omega_0$.

In order to make this discussion more quantitative, 
we have evaluated numerically the spectral weight of 
the different band states, 
\begin{equation}
Z_{\vec{K}}=\frac{\left|
\langle {\mit \Psi}_{0,\vec{K}}^{(N-1)}
|\tilde{c}^{}_{\vec{Q}-\vec{K},\sigma}| {\mit \Psi}_{0,\vec{Q}}^{(N)}\rangle
\right|^2}{
\left|\langle {\mit \Psi}_{0,\vec{Q}}^{(N)}
|\tilde{c}^{\dagger}_{\vec{Q}-\vec{K},\sigma}\tilde{c}^{}_{\vec{Q}-\vec{K},
\sigma}
| {\mit \Psi}_{0,\vec{Q}}^{(N)}\rangle\right|^2}\,,
\label{zk}
\end{equation}
where $ |{\mit \Psi}_{0,\vec{Q}}^{(N-1)}\rangle$ denotes
the one--hole (polaron) state being lowest in energy within the 
$\vec{Q}$--sector. 
The results for $Z_{\vec{K}}$ are displayed  in 
Fig.~4~(b). As already discussed in Sec.~III, 
$Z_{\vec{K}}$ is strongly reduced from unity in the pure
t--J model (but seems to remain finite 
in the thermodynamic limit if $J>0$).
Including a weak adiabatic EP interaction ($\ep=0.2$, $\omega_0=0.2$),
the band states with energies larger than  $E_0+\omega_0$ change
their character from ``hole--like'' to ``phonon--like''.
Consequently, we observe a nearly complete suppression of 
$Z_{(0,0)}$ and $Z_{(\pi,\pi/2)}$. Increasing the phonon frequency, 
this tendency is reversed (see the curve for $\omega_0=0.8$). 
Note that similar effects have been discussed quite recently 
in terms of the single--electron Holstein model~\cite{St96,WF97}. 

To complete this section, we show in Fig.~5 the variation of 
$Z_{\vec{K}}$ with the EP coupling strength for the $\vec{K}$--vectors
of the ten--site lattice studied in Sec.~III. 
The wave--function renormalization factor $Z_{\vec{K}}$
can be taken as a measure of the ``contribution'' of the hole 
(dressed at $\ep=0$ by spin--wave excitations) to the polaronic QP (having
total momentum $\vec{K}$). The data obtained at weak EP 
coupling unambiguously confirm the different nature of band states 
in this regime: we found practically zero--phonon ``hole'' states 
at the band minima ($\vec{K}=(3\pi/5,\pi/5)$, triangles down) 
and ``phonon'' states, which are only weakly affected by the hole, 
around the (flat) band maxima ($\vec{K}=(\pi,\pi)$, triangles up).   
With increasing $\ep$, a strong ``mixing'' of holes and phonons 
takes place, whereby both quantum objects completely loose their 
own identity. Concomitantly $Z_{\vec{K}}$ decreases for the ``hole--like''
states but increases (first of all) for the ``phonon--like'' states.   
At large $\ep$, a small lattice hole polaron is formed, which, 
according to the numerical results, 
has an extremely small spectral weight. Hence the question arises 
whether the lattice hole polaron is a ``good'' QP in the sense that
one can construct a QP operator,
$\tilde{c}_{\vec{K}\sigma}\to\tilde{d}_{\vec{K}\sigma}$, 
having large spectral weight at the lowest pole in the spectrum.  
Indeed, it was demonstrated in recent ED work that it is possible
to construct such a composite electron/hole--phonon (polaron) operator 
by an appropriate phonon dressing of $\tilde{c}_{\vec{K}\sigma}$ 
for the Holstein model~\cite{FLW97} as well as for the t--J model 
coupled to buckling/breathing modes~\cite{SPS97}. 
\section{Optical conductivity}
Using standard linear response theory, the Kubo formula for 
the real (absorbent) part of the frequency--dependent optical
conductivity gives two physically distinct contributions~\cite{SS90,Ps96}:    
\begin{equation}
\Re\mbox{e}\, \sigma_{xx}(\omega ) = {\cal D}\delta(\omega) +
\sigma^{reg}_{xx}(\omega)\,,\qquad \omega\geq 0\,. 
\label{resig}
\end{equation}
The first so--called Drude term at $\omega=0$ is 
due to the free acceleration of the charge carriers by the electric 
field and the second term, frequently called the ``regular term'',
is due to finite frequency  dissipative optical transitions 
to excited quasiparticle states. More explicitly, the regular part 
can be written in spectral representation at $T=0$ as~\cite{Da94} 
\begin{equation}
\sigma^{reg}_{xx}(\omega)=\frac{e^2\pi}{N}\sum_{n\neq 0}
\frac{\left|\langle {\mit \Psi}_{n,\vec{K}}^{(N-1)} | \hat{j}_x^{(p)} |  
{\mit \Psi}_{0,\vec{K}}^{(N-1)}\rangle \right|^2}
{E^{(N-1)}_{n,\vec{K}}-E^{(N-1)}_{0,\vec{K}}} \,\;\delta
\left[\omega-(E_{n,\vec{K}}^{(N-1)}-E_{0,\vec{K}}^{(N-1)})\right]\,,
\label{sigreg}
\end{equation}
where the summation is taken over the complete set of eigenstates
with excitation energies 
$\omega=[E_{n,\vec{K}}^{(N-1)}-E_{0,\vec{K}}^{(N-1)}]$ 
in the one--hole ($N-1$ electron) subspace.  
For the HtJM the (paramagnetic) current density operator $j_x^{(p)}$ 
has the form  
\begin{equation}
\label{stromop}
j_x^{(p)} =it\sum_{i\sigma}( \tilde{c}_{i,\sigma}^{\dagger}
\tilde{c}_{i+x,\sigma}^{} - \tilde{c}_{i+x,\sigma}^{\dagger}
\tilde{c}_{i,\sigma}^{})\,.
\end{equation}
Actually in~(\ref{sigreg}) an optical transition can take place only 
within the $\vec{K}$--sector of the ground state.

The expression~(\ref{sigreg}) has been used to calculate 
$\sigma^{reg}(\omega)$ numerically for the t--J 
(e.g., see Ref.~\cite{Da94}) and Holstein models~\cite{AKR94b,CSG97,FLW97} 
on finite lattices employing ED techniques. 
First results, obtained recently for the 1D
HtJM~\cite{Fer94}, indicate that EP coupling effects 
may be at least partly responsible for the experimentally observed 
frequency dependence of the optical conductivity in the $\rm CuO_3$ chains 
of the 1:2:3 family of high--$T_c$ oxides.   

In Fig.~6 we show $\sigma^{reg}(\omega)$ of the 2D ten--site HtJM,
determined from~(\ref{sigreg}) with at most $M=15$ phonons, 
for typical values of the hole--phonon coupling and $J=0.4$.
In the weak EP coupling regime and for phonon frequencies
$\omega_0 \stackrel {>}{\sim}{\mit \Delta}E^{t-J}$, we recover the main
features of the optical absorption spectrum of the 2D t--J model~\cite{Po91}, 
i.e., an ``anomalous'' broad mid--infrared 
band $[J\stackrel{<}{\sim}\omega\stackrel{<}{\sim}2t]$, separated from 
the Drude peak [${\cal D}\delta (\omega)$; not shown] by a ``pseudo--gap''
$\simeq J$, and an ``incoherent'' tail up to $\omega\simeq 7t$
(compare panels (a) and (b)).
Note particularly the even quantitative agreement between the
$\omega$--integrated spectral weight functions
\begin{equation}
S^{reg}(\omega)=\int_0^{\omega} d\omega^{\prime} 
\,\sigma_{xx}^{reg}(\omega^{\prime})
\label{sreg}
\end{equation}
depicted in Figs.~6~(a) and~(b). Since for the parameters used in (b),
$\ep=0.1$ and $\omega_0=0.8$, the averaged kinetic energy per site 
$\langle - \cH_t/N \rangle$ is only weakly renormalized by the EP interaction
(cf. Fig.~10 of Ref.~\cite{WRF96}; $\cH_t$ denotes the first term
in~(\ref{htjm})), also the Drude weight 
\begin{equation}
{\cal D}=\frac{\pi e^2}{2N} \langle- \cH_t \rangle - 2 S^{reg}(\infty)\,,
\label{druwe}
\end{equation}
obtained via the f--sum rule, should approximately be the same 
as for the non--interacting system ($\ep=0$). 
As discussed in the  preceding section, 
the coherent band structure gets stronger renormalized at large $\ep$, 
i.e., the QP band states have less spectral weight.
At the same time the phonon distribution function
in the ground state,  $|c^{m}_0|^2$ (see Eq.~(18) Appendix), 
becomes considerably broadened~\cite{WRF96,FWBB97}, and the position of
its maximum is shifted to larger values of $m$. Consequently the 
overlap with excited multi--phonon states is enlarged and 
the optical response is enhanced at higher energies. 
This redistribution of spectral weight from low to high energies
can be seen in Fig.~6~(c). As expected the transition
to the lattice hole--polaron state, at about 
$\varepsilon_{p}^c(J=0.4, \omega_0=0.8)\simeq 2.0$, 
is accompanied by the development of a broad
maximum in $\sigma^{reg}(\omega)$, whereas the Drude weight as well as 
the low--frequency optical response become strongly suppressed.
Contrary, in the strong--coupling anti--adiabatic regime ($\ep=4.0,\;
\omega_0=10.0$; see panel~(d)), 
the low--frequency part of $\sigma^{reg}(\omega)$ is much 
less affected, but the optical spectrum shows additional
superstructures corresponding to ``interband'' transitions between
t--J--like absorption bands with different numbers of phonons~\cite{FWBS97}.
  
Let us now make contact with the experimentally observed characteristics
of the mid--infrared spectra in the doped perovskites (for an review 
of the experimental and theoretical work on the optical conductivity see 
Refs.~\cite{TT92,Da94,AKR94b,SAL95}). Needless to say, that 
it is out of the scope of our ten--site cluster diagonalization study to give 
a quantitative theoretical description of the complex optical properties of
particular copper and nickel oxides. But we would like to stress  
that the 2D HtJM seems to contain the key ingredients 
to reproduce, at least qualitatively, the principal features of the optical 
absorption spectra of these compounds. This can be seen by comparing
the results shown in Figs.~6 (b) and (c), which correspond to the weak and
strong EP coupling situations realized in the 
cuprate ($\rm La_{2-x}Sr_xCuO_4$) and nickelate 
($\rm La_{2-x}Sr_xNiO_4$) systems, respectively.
The ``effective'' EP interaction $\ep/t$ in the nickelates 
is estimated to be about one order of magnitude larger than in 
the cuprates simply because of the much smaller transfer
amplitude ($t\simeq 0.08$ eV~\cite{BE93}). 
According to the internal structure of the low--spin state,
the hopping transport of spin--1/2 composite holes in a 
spin--1 background is rather complex; implying,
within an effective single--band description, a strong reduction
of the transition matrix elements~\cite{LF97}.
A striking feature of the absorption spectra in the cuprate
superconductors is the presence of a broad mid--infrared (MIR) band,
centered at about 0.5~eV in lightly doped $\rm La_{2-x}Sr_xCuO_4$
(which, using $t\sim 0.3$~eV, means that $\omega\sim 1.5$).   
Such a strong MIR absorption is clearly observed in Fig.~6~(b),
which refers to  the weak EP coupling case. 
Since it also appears in the pure t--J model
(cf. panel (a)), this MIR band seems to be caused by the spin
fluctuations around the charge carrier~\cite{Da94}. Obviously, 
it is quite difficult to distinguish the spectral weight, produced 
by the dressing of the hole due to the ``bag'' of reduced 
antiferromagnetism in its neighbourhood~\cite{SWZ88}, 
from other (e.g. hole--phonon coupling) processes 
that may contribute to the MIR band observed experimentally.
The results presented for the HtJM in Fig.~6~(b) support 
the claims, however,  that the MIR band in the cuprates has a
mainly ``electronic'' origin, i.e., the lattice polaron effects are 
rather weak. The opposite is true for their isostructural
counterpart, the nickelate system, where the MIR absorption band 
has been ascribed by many investigators to ``polaronic''
origin~\cite{BE93,Caea97}. Within the HtJM such a situation
can be modeled by the parameter set used in Fig.~6~(c).
If we fix the energy scale by  $t\sim 0.08$~eV (which is
the estimate for the nearest neighbour transfer integral
in $\rm La_{2-x}Sr_xNiO_4$ given in Ref.~\cite{BE93}),  
the maximum in the optical absorption is again located at about 0.5~eV.
The whole spectrum clearly shows lattice polaron characteristics, where
it seems that the lattice hole polarons are of small--to--intermediate 
size~\cite{Emi95}.
Most notably, we are able to reproduce the experimentally observed asymmetry
in the shape of the spectrum, in particular the very gradual decay 
of $\sigma^{reg}(\omega)$ at high energies. 
It is worth mentioning that this behaviour cannot 
be obtained from a simple fit to the analytical expressions
derived for the small polaron hopping conductivity~\cite{Re67,SPH91,BE93}.  
Exploiting the f--sum rule we found that there are almost no
contributions from band--like carriers in agreement with the
experimental findings~\cite{BE93,Caea97}.

Finally, we would like to consider the case $J=0.1$, 
which certainly has no relevance with respect to the perovskite 
materials but this parameter regime is interesting from a theoretical 
point of view. For $\ep=0$ (t--J model), the momentum of the Nagaoka ground 
state is $\vec{K}=(0,0)$ and the hole can propagate 
without any spin scattering in the ferromagnetic background. 
Therefore all the spectral weight stays in the Drude pole at $\omega =0$  
and the regular (incoherent) part of the optical conductivity 
vanishes exactly.  In principle,
at arbitrary small EP interactions an optical transition can
be achieved by ``adding'' phonons with momentum ($-\vec{K}$)
to band states $E_{\vec{K}}$ (in order to reach the $\vec{K}=(0,0)$ 
sector). However, the overlap of these excited states with the 
almost zero--phonon ground state is extremely small and,   
as a direct consequence, $S^{reg}(\infty)$ is negligible  
(see Fig.~7~(a)). A completely different situation arises when
the charge carrier becomes ``self--trapped'' due to the lattice
distortions. When the lattice hole--polaron formation takes place, 
the momentum of the ground state of the 2D ten--site HtJM 
is switched from (0,0) to $\vec{K}=(3\pi/5,\pi/5)$ 
and, what is more important, the AFM exchange interaction 
becomes more effective. This gives rise to the pronounced
low--frequency peak structures found in Fig.~7~(b), which are  
originated from spin excitations on an energy scale of the order of
$J$. At higher energies we recover the characteristic features
of the lattice hole--polaron spectrum discussed in Fig.~6~(c).  
\section{Summary} 
We have investigated the spectral properties of the two--dimensional 
Holstein t--J model as a generic model for studying polaronic effects
in systems with strong Coulomb correlations. The use of purely
numerical techniques allows to treat the electron and phonon 
degrees of freedom on an equal footing. In the numerical work 
the full dynamics of all the phonon modes was taken into account on finite 
lattices with up to 16 sites. The efficiency and accuracy of the
employed phonon Hilbert space truncation, Lanczos diagonalization, 
Chebyshev recursion and maximum entropy methods have been demonstrated 
in the Appendix. 

In conclusion, we have shown that an additional
hole--phonon coupling strongly affects the nature of 
the quasiparticle excitations and the charge--carrier
transport in the 2D t--J model. Our results indicate that
the magnetic prelocalization of the holes, due to 
strong spin correlations, strengthens the effect of 
the hole--phonon interaction. Thus the tendency
towards lattice (hole) polaron formation is enhanced in strongly
correlated electron systems.  

The wave--vector resolved single hole spectral function 
yields a flattening of the coherent band dispersion 
in the vicinity of the band maxima for the adiabatic 
weak and intermediate hole--phonon coupling cases.
As the hole--phonon interaction increases, a narrow lattice
hole--polaron band separates from the rest of the spectrum but 
the spectral weight of these quasiparticle band states
becomes more and more suppressed.  In the anti--adiabatic 
regime the low--energy excitations of the t--J model 
are less affected by the hole--phonon interaction;
at higher energies we found ``interband'' transitions 
corresponding to multi--phonon absorption processes. 

The formation of lattice hole polarons is accompanied by significant
changes in the optical response of the system.  
Both the coherent Drude part and the low--frequency regular contribution
to the optical conductivity are strongly reduced and, concomitantly, 
we observe a substantial spectral weight transfer to higher energies.
In the intermediate--to--strong hole--phonon coupling regimes, i.e.
in the vicinity of the lattice hole--polaron transition, 
the lineshape of the optical absorption spectra 
is highly asymmetric with a long incoherent tail 
at high frequencies, which also seems to be in accordance 
with recent mid-- and far--infrared optical experiments on 
$\rm La_{2-x}Sr_xNiO_4$~\cite{BE93,Caea97}. The present
calculations point toward the importance of strong spin correlation 
and polaronic effects in explaining the origin of the anomalous 
mid--infrared absorption band observed 
in the cuprate and nickelate compounds, respectively.
Of course, for a realistic modelling of the complex
high--spin low--spin nickelate system, one has to 
take into account more complicated hopping and spin 
exchange processes. Work along this line is in progress.
\acknowledgments
This work was performed  under the auspices of Deutsche
Forschungsgemeinschaft, SFB 279.
Special thanks go to the LRZ M\"unchen, the HLRZ J\"ulich and 
HLR Stuttgart for the generous granting of their parallel computer
facilities. We have benefited from discussions with A. Basermann,
A. R. Bishop,  H. B\"uttner, D. Ihle, J. Loos, and H. R\"oder.  
We are particularly indebted to R.~N.~Silver for putting his 
maximum entropy code at our disposal.
\section*{Appendix: Numerical methods}
\subsection*{Symmetrized basis and Hilbert space truncation}
{\sl 1. Basis symmetrization.} 
The total Hilbert space of the HtJM can be written as the tensorial
product space of electrons and phonons, spanned by the
complete basis set  
\begin{equation}
\left\{|{\mit \Phi}_{uv}\rangle=
|u\rangle_{el} \otimes |v\rangle_{ph} \right\}
\label{basis1}
\end{equation}
with
\begin{eqnarray}
|u\rangle_{el} &=& \prod_{i=1}^N \prod_{\sigma = \uparrow,\downarrow} 
(\tilde{c}_{i\sigma}^{\dagger})^{n_{i\sigma,u}}\;\; 
|0\rangle_{el},\qquad\quad\; n_{i\sigma,u} \in \{0,1\}\,,\\
|v\rangle_{ph} &=& \prod_{i=1}^N \frac{1}{\sqrt{m_{i,v} !}} 
(b_i^{\dagger})^{m_{i,v}}\; |0\rangle_{ph},\quad m_{i,v} \in 
\{0,\ldots,\infty\}\,.
\end{eqnarray}
Here $u=1,\ldots,D_{el}$ and $v=1,\ldots,D_{ph}$ label the
basic states of the electronic and phononic subspaces with
dimensions $D_{el}={N \choose N_{\sigma}}{N -N_{\sigma}\choose N_{-\sigma}}$
and $D_{ph}=\infty$, respectively.

Since the Hamiltonian~(1) commutes with the electron number operator
${\cal N}_{el}= \sum_{i=1}^N(\tilde{n}_{i,\uparrow}+\tilde{n}_{i,\downarrow})$ 
and the $z$--component of the total spin ${\cal S}^{z}=
\frac{1}{2}\sum_{i=1}^N(\tilde{n}_{i,\uparrow}-\tilde{n}_{i,\downarrow})$ 
the basis~(\ref{basis1}) has been constructed for    
fixed $N_h=N-N_{el}$ and $S^z=S^z_{min}$.
To further reduce the dimension of the total Hilbert space,
we can exploit the space group symmetries 
[translations ($G_T$) and point group operations ($G_L$)] 
and the spin--flip invariance [($G_S$); $S^z=0$--subspace only].
Clearly, working on finite bipartite clusters ($N=k^2+l^2$, $k$ and $l$ 
are both even or odd integers) with periodic boundary conditions,
we do not have all the symmetry properties of the 2D square lattice. 
Restricting ourselves to the one--dimensional
non--equivalent irreducible representations of the group 
$G(\vec{K})=G_T\times G_L(\vec{K})\times G_S$, 
we can use the projection operator
\begin{equation}
{\cal P}_{\vec{K},rs}=\frac{1}{g(\vec{K})} 
\sum_{{\cal G} \in G(\vec{K})}\chi_{\vec{K},rs}^{(\cal G)}\;{\cal G} 
\label{pop}
\end{equation}
(with $[{\cal H},{\cal P}_{\vec{K},rs}]=0$, 
${\cal P}_{\vec{K},rs}^{\dagger}={\cal P}_{\vec{K},rs}$ and
${\cal P}_{\vec{K},rs}\;{\cal P}_{\vec{K}^{\prime},r^{\prime}s^{\prime}}
={\cal P}_{\vec{K},rs}\;  
\delta_{\vec{K},\vec{K}^{\prime}}\;\delta_{r,r^{\prime}}\;
\delta_{s,s^{\prime}}$),
in order to generate a new symmetrized basis set:
$\{|{\mit \Phi}_{uv}\rangle\} \stackrel{\cal P}{\to}  
\{|\tilde{\mit \Phi}_{\tilde{u}v}\rangle\}$. 
In (\ref{pop}), ${\cal G}$ denotes the  $g(\vec{K})$ elements of 
the group $G(\vec{K})$ and $\chi_{\vec{K},rs}^{(\cal G)}$ 
is the (complex) character of 
${\cal G}$ in the $[\vec{K},rs]$--representation, where  
$\vec{K}$ refers to one of the $N$ allowed wave vectors in the 
first Brillouin zone,  $r$ labels the irreducible representations 
of the little group of $\vec{K}$, $G_L(\vec{K})$, and $s$
parameterizes $G_S$.   
For an efficient parallel implementation of the 
matrix vector multiplications (MVM)
required diagonalizing the Hamiltonian, it is extremely important,  
that the symmetrized basis can be constructed preserving the tensor 
product structure of the Hilbert space:
\begin{equation}
\{|\tilde{\mit \Phi}_{\tilde{u}v}\rangle=
N^{[\vec{K}rs]}_{\tilde{u}v} {\cal P}_{\vec{K},rs}
\left[ |\tilde{u}\rangle_{el}
\otimes |v\rangle_{ph}\right]\}\,. 
\end{equation}
Here $\tilde{u}=1,\ldots, \tilde{D}_{el}^{g(\vec{K})}$ 
$[\tilde{D}_{el}^{g(\vec{K})}\sim D_{el}/g(\vec{K})]$, and 
the $N^{[\vec{K}rs]}$ are normalization factors. 

{\sl 2. Phonon Hilbert space truncation.}
Since the Hilbert space associated to the phonons is infinite
even for a finite system, we apply a truncation procedure~\cite{II90,WRF96}
retaining only basis states with at most $M$ phonons:  
\begin{equation}
\{ |v\rangle_{ph}\; ; \;m_v=\sum^N_{i=1} m_{i,v} \le M \}.
\end{equation}
The resulting Hilbert space has a total dimension 
$\tilde{D}^M=\tilde{D}_{el}^{g(\vec{K})}\times D_{ph}^M$ with 
$D_{ph}^M =\frac{(M+N)!}{M!N!}$, and a general state 
of the HtJM is represented as   
\begin{equation}
|{\mit \Psi}_{\vec{K},rs}\rangle = 
\sum_{\tilde{u}=1}^{\tilde{D}_{el}^{g(\vec{K})}} 
\sum_{v=1}^{D^M_{ph}} \tilde{c}_{\tilde{u}v}\,
|\tilde{\mit \Phi}_{\tilde{u}v}\rangle\,. 
\end{equation}
It is worthwhile to point out that, switching from a real space representation
to a momentum space description, our truncation scheme takes 
into account {\it all} dynamical phonon modes. This has to be 
contrasted with the frequently used single--mode 
approaches~\cite{PSSH96,AP98}. 
In other words, depending on the model parameters and the band filling, 
the system ``decides'' by itself how the $M$ phonons will be 
distributed among the independent Einstein oscillators related 
to the $N$ Wannier sites or, alternatively, among the $N$ different 
phonon modes in $\vec{Q}$--space. Hence with the 
same accuracy phonon dynamical effects on lattice distortions 
being quasi--localized in real space 
(such as polarons, Frenkel excitons,\ldots) or in momentum space 
(like polaronic superlattices, charge--density--waves,\ldots)
can be studied. 

Of course, one has carefully to check for the convergence of
the above truncation procedure by calculating the ground--state
energy as a function of the cut--off parameter $M$. 
In the numerical work convergence is assumed to be achieved if 
$E_0$ is determined with a relative error
\begin{equation}
\Delta E_0^{(M)} = \frac{E_0(M)-E_0(M-1)}{E_0(M)}\leq 10^{-6}\,.
\end{equation}
In addition we guarantee that the phonon distribution function 
\begin{equation}
|c^{(m)}|^2(M) = \sum^{\tilde{D}^{g(\vec{K})}_{el}}_{\tilde{u}=1} 
\sum^{D^M_{ph}}_{\stackrel{{\displaystyle {}_{v=1}}}{\{m_v=m\}}} 
|\tilde{c}_{\tilde{u}v}|^2\,,
\end{equation}
which gives the different weights of the $m$--phonon states in the 
ground--state $|{\mit \Psi}_0\rangle$, becomes independent of $M$ 
and $|c^{(M)}|^2(M)\leq 10^{-6}$. 

To illustrate the $M$ dependences of the phonon distribution function
and the ground--state energy, we have shown both quantities in Fig.~8 
for the HtJM with $\ep=4.0$, $\hbar\om =0.8$, and $J=0.4$.
Fig.~8 proves that our truncation procedure is very well controlled
even in the strong EP coupling regime, where multi--phonon states become
increasingly important. Although not mentioned each time, all 
the results presented in this paper have been obtained
with a sufficiently large value of $M$ ensuring the 
above convergence criteria for both static (ground state) and 
dynamic (excited state) properties.

Finally we would like to stress that it is not possible
to keep all non--vanishing matrix elements of our ten--site Holstein t--J 
Hamiltonian in the memory of the present day parallel computers 
because,  in spite of applying the symmetrization and Hilbert space 
truncation procedures, the total dimension of the problem 
is still very large.
Therefore, performing the matrix--vector--multiply
operations,  we create the non--zero matrix elements 
of the Hamiltonian simultaneously. 
For the example considered in Fig.~8,
we have $\tilde{D}^M=4.12\times 10^{8}$ ($M=15$) phonons, 
and each MVM takes about 100 seconds cpu time on a 
CRAY T3E with 126 processors.  
\subsection*{Algorithms for estimating spectral functions} 
The numerical calculation of spectral functions, 
\begin{eqnarray}
A^{\cal O}(\omega)&=&-\lim_{\varepsilon\to 0^+}\frac{1}{\pi} \Im m \left[
\langle{\mit \Psi}_{0}
|{\bf O}^{\dagger}\frac{1}{\omega - {\bf H} +E_0 +i\varepsilon}{\bf O}^{} 
|{\mit \Psi}_{0}\rangle\right]\nonumber\\&=&
\sum_{n=0}^{D-1}|\langle{\mit \Psi}_{n}|{\bf O}^{\dagger}|
{\mit \Psi}_{0}\rangle |^{2}\delta [\omega - (E_{n} - E_{0})]\,,
\end{eqnarray}
where ${\bf O}$ is the matrix representation of a certain 
operator ${\cal O}$ [e.g., the destruction operator
$\tilde{c}_{\vec{K},\sigma}$ of an electron with  
momentum $\vec{K}$ and spin $\sigma$ 
if one wants to calculate the single hole spectral
function~(2)], involves very large sparse Hamilton matrices ${\bf H}$.
For example, for the EP systems studied in this paper,  
we have calculated spectral properties of the HtJM acting  
in Hilbert spaces with total dimensions up to $D=10^9$. 
Finding all ($D$) eigenvectors and eigenstates of such  huge 
Hamiltonian matrices is impossible, because the cpu time required for
exact diagonalization of ${\bf H}$ scales as $D^3$ and memory as $D^2$. 
Fortunately, there exist very accurate and well--conditioned 
linear scaling algorithms for a direct approximate calculation of 
$A^{\cal O}(\omega)$. 
 
In this appendix we compare the widely used
{\it spectral decoding method} (SDM)~\cite{ZSP94}, which is a
refinement of the standard  
{\it Lanczos recursion method} (LRM)~\cite{GB87},
with two more recent Chebyshev recursion techniques,  
the {\it kernel polynomial method} (KPM)~\cite{SR94} and the  
{\it maximum entropy method} (MEM)~\cite{SR97}. 
To come to the conclusion first, the MEM is found to be the most 
efficient and numerically stable method for high--energy 
resolution applications such as the lattice polaron formation problem. 

In the following we briefly review the basics of the 
SDM, KPM and MEM (for a more detailed discussion see, 
e.g., Refs.~\cite{Da94,SR94,SR97,Ba97}), 
and illustrate their efficiency and accuracy with several examples.

{\sl 1. SDM.}
First one calculates by iterative 
matrix--on--vector multiplications, 
\begin{equation}
|\bar{\mit \Phi}_{l+1}\rangle=
{\bf H}^{D}|\bar{\mit \Phi}_{l}\rangle
-\frac{\langle\bar{\mit \Phi}_{l}|{\bf H}^{D}
|\bar{\mit \Phi}_{l}\rangle}{\langle\bar{\mit \Phi}_{l}|
\bar{\mit \Phi}_{l}\rangle}|\bar{\mit \Phi}_{l}\rangle
-\frac{\langle\bar{\mit \Phi}_{l}|
\bar{\mit \Phi}_{l}\rangle}{\langle\bar{\mit \Phi}_{l-1}|
\bar{\mit \Phi}_{l-1}\rangle}|\bar{\mit \Phi}_{l-1}\rangle\,,
\label{lr}
\end{equation} 
a {\it tridiagonal} matrix  $[{\bf T}^L]_{l,l^{\prime}}=
\langle\bar{\mit \Phi}_{l}|{\bf H}^{D}|
\bar{\mit \Phi}_{l^{\prime}}\rangle$ 
with dimension $L\ll D $, starting out from 
$|\bar{\mit \Phi}_1\rangle={\bf H}^{D}|\bar{\mit \Phi}_{0}\rangle
-\frac{\langle\bar{\mit \Phi}_{0}|{\bf H}^{D}
|\bar{\mit \Phi}_{0}\rangle}{\langle\bar{\mit \Phi}_{0}|
\bar{\mit \Phi}_{0}\rangle}|\bar{\mit \Phi}_{0}\rangle$,  
where ${\bf H}^D$ denotes the Hamilton matrix acting in a $D$--dimensional
Hilbert space and $|\bar{\mit \Phi}_{0}\rangle$ is an arbitrary initial state 
having finite overlap with the true ground state $|{\mit \Psi}_{0}\rangle$. 
Applying the Lanczos recursion~(\ref{lr}), the eigenvalues $E_n$ and 
eigenvectors $|{\mit \Psi}_{n}\rangle $ of ${\bf H}^D$ 
are approximated by those of  ${\bf T}^L$~\cite{CW85},  $E_n^L$ and 
$|{\mit \Psi}^L_{n}\rangle=\sum_{l=0}^{L-1}c_{n,l}^L|\bar{\mit \Phi}_{l}
\rangle $, respectively.  

Then, having determined the ground state $|{\mit \Psi}_{0}^L\rangle$ by the 
Lanczos technique, we can use again the recursion relation~(\ref{lr}),
but with the initial state
\begin{equation}
|\bar{\mit \Phi}_0\rangle=\frac{{\bf O}|{\mit \Psi}^L_{0}\rangle}{\sqrt{ 
\langle {\mit \Psi}^L_{0}|{\bf O}^{\dagger}{\bf O}|{\mit \Psi}^L_{0}
\rangle}}\,,
\label{init}
\end{equation}
to determine within the {\it SDM} an approximative spectral function,
\begin{equation}
\bar{A}^{\cal O}(\omega)=\sum_{n=0}^{L-1} |c_{n,0}^L|^2 
\langle{\mit \Psi}_{0}|{\bf O}^{\dagger}{\bf O}|{\mit \Psi}_{0}\rangle\,
\delta[\omega -(E_n^L-E_0^L)]\,, 
\label{asdm}
\end{equation} 
built up by $L$ $\delta$-peaks. Of course,
the true spectral function $A^{\cal O}(\omega)$ has  $D$
$\delta$-peaks. According to the  Lanczos phenomenon the approximated 
spectral weights and positions of the peaks converge to their true values 
with increasing $L$. Some of the main problems of the SDM/LRM are:
(i) The convergence is not uniform in the whole energy range.  
(ii) There exist so--called spurious peaks, which appear 
and disappear as $L$ is increased, i.e., when the iteration proceeds.
(iii) Without computationally expensive re--orthogonalization only a 
few hundred iterations are possible. 

{\sl 2. KPM and MEM}. The idea behind the Chebyshev recursion methods is 
to expand the $\delta$--function contained in $A(\omega)$ in a series 
of Chebyshev polynomials $T_m(x)$,
\begin{equation}
A^{\cal O}(x)=\frac{1}{\pi
\sqrt{1-x^{2}}}\left(\mu_{0}^{\cal O}+2\sum_{m=1}^{\infty}\mu_{m}^{\cal O}T_{m}(x)\right)\,,
\label{akpm}
\end{equation}
with the coefficients (moments) 
\begin{equation}
\mu_m^{\cal O}=\int_{-1}^{1}dx\,T_{m}(x)A^{\cal O}(x)=\langle {\bf O}^{\dagger}
{\mit \Psi}_{0}|T_{m}({\bf X})|{\bf O}^{\dagger}{\mit \Psi}_{0}\rangle\,,
\label{mkpm}
\end{equation}
$x=(\omega -b)/a$, $a=(E_{max}-E_{min})/2(1-\epsilon)$, and 
$b=(E_{max}+E_{min})/2$. Since the Chebyshev polynomials are defined 
on the interval $[-1,1]$, a re--scaled Hamiltonian 
${\bf X}=({\bf H}-b)/a$ has to be considered, whose eigenvalues $x_n$ are 
in the range $[-(1-\epsilon),1-\epsilon]$. Here a small 
constant $\epsilon$ is introduced in order to avoid
convergence problems at the endpoints of the interval 
(a typical choice is $\epsilon \sim 0.01$ which has only
1\% impact on the energy resolution~\cite{SR97}). 
The $\mu_m^{\cal O}$ can be easily computed by iterative MVM. 
Truncating the expansion~(\ref{akpm}) at $\mu_{\tilde{M}}^{\cal O}$, 
the main difference between KPM and MEM is the way 
the spectral function is reconstructed from the  
remaining $\tilde{M}$ moments.

The KPM is a {\it linear} Chebyshev approximation to the spectrum,
$\tilde{A}^{\cal O}(x)\simeq A^{\cal O}(x)$, 
using $\tilde{M}$ moments only.   
Here the abrupt truncation of the series produces the well--known
Gibbs oscillations which, however, can be minimized by multiplying 
the moments by appropriate damping factors~\cite{SRVK96}.

On the other hand, the MEM is a {\it non--linear} optimization procedure   
for calculating from the $\hat{M}$ known moments a representation of a  
(reconstructed) spectral function, 
$\hat{A}^{\cal O}(\phi)=\check{A}^{\cal O}(x) \sin \phi$ with $x=\cos \phi$, 
that corresponds to a Chebyshev series with 
$ \hat{M}^{eff}=F\cdot \hat{M}$ summands. 
The $ \hat{M}^{eff}$ moments
\begin{equation}
\hat{\mu}_m^{\cal O}=\int_0^{\pi} d\phi \cos(m\phi)\hat{A}^{\cal O}(\phi)
\label{mmem}
\end{equation}  
are determined by maximizing the information theoretical relative
entropy~\cite{SR97}
\begin{equation}
{\cal S}=\int_0^{\phi}d\phi\left[\hat{A}^{\cal O}(\phi)-\hat{A}_0^{\cal O}(\phi)-
 \hat{A}^{\cal O}(\phi)\ln \left(\frac{\hat{A}^{\cal O}(\phi)}{\hat{A}_0^{\cal O}(\phi)}\right)\right]
\label{smem}
\end{equation}
subject to the constraint that the first $\hat{M}$ moments 
of $\hat{A}^{\cal O}(\phi)$ are equal to the $\hat{M}$ known moments. 
$\hat{A}_0^{\cal O}(\phi)$ is a default model for $\hat{A}^{\cal O}(\phi)$.
We choose the KPM result: $\check{A}_0^{\cal O}(x)=\tilde{A}^{\cal O}(x)$. 
$F$ is typically $8\ldots 32$.

For the MEM (KPM) the calculation of the Chebyshev moments and not
the reconstruction of the spectral function requires the essential
part of cpu time and memory. Clearly this is because in moment production
MVM with ${\bf H}^D$ is involved,
whereas the  reconstruction  of the spectrum only processes 
the $\tilde{M}$ $(\hat{M})$ moments. 

In the numerical work we used the following algorithm 
producing two moments per MVM with ${\bf H}^D$:
(i) Determine from Lanczos diagonalization the ground state  
$|{\mit \Psi}_{0}\rangle$ and the extreme eigenvalues $E_{min}=E_0$ 
and $E_{max}$  of ${\bf H}^D$. 
(ii) Compute the starting conditions  
\begin{eqnarray}
   |\hat{\mit \Phi}_{0}\rangle&=&{\bf O}^{\dagger}|{\mit \Psi}_{0}\rangle, 
   \qquad|\hat{\mit \Phi}_{1}\rangle={\bf X}|\hat{\mit \Phi}_{0}\rangle\,,\\ 
  \mu^{\cal O}_{0}&=& \langle\hat{\mit \Phi}_{0}|\hat{\mit \Phi}_{0}\rangle , 
  \qquad \mu_{1}^{\cal O}=\langle
\hat{\mit \Phi}_{0}|\hat{\mit \Phi}_{1}\rangle\,.
\label{memit1}
\end{eqnarray}
(iii) Use the recurrence relations for Chebyshev polynomials to
calculate the $\hat{M}$ moments according to
\begin{eqnarray}
  |\hat{\mit \Phi}_{m}\rangle  & = &
  2{\bf X}|\hat{\mit \Phi}_{m-1}\rangle - |\hat{\mit \Phi}_{m-2}\rangle\,, \\ 
  \mu_{2m}^{\cal O} & = & 2
    \langle\hat{\mit \Phi}_{m}|
    \hat{\mit \Phi}_{m}\rangle-\mu_{0}^{\cal O}\,,\\ 
   \mu_{2m-1}^{\cal O}& = & 2
   \langle\hat{\mit \Phi}_{m}|\hat{\mit \Phi}_{m-1}\rangle-\mu_{1}^{\cal O}\,.
\label{memit2}
\end{eqnarray}
Note that applying this procedure only two $D$--dimensional vectors 
have to be stored. The algorithm is numerically 
very stable, so in contrast to the SDM/LRM  (continued fraction
based technique~\cite{Da94}) thousands of Chebyshev moments
can be calculated with satisfying accuracy.

{\sl 3. Examples}. Figure~9 demonstrates the non--uniform
convergence of the SDM/LRM for the dynamical pair correlation spectrum 
of the t--J model (for a discussion of the physics described by
$A^{\mit \Delta}(\omega)$ see, e.g., Refs.~\cite{Da94,WBF98}). 
Obviously, we found a much faster
convergence in the low--energy regime. Increasing $L$ the 
peak--resolution becomes better at higher energies as well, but now one
is confronted with the re--orthogonalization problem mentioned above. 

Another typical problem concerns the proper description of gap structures.
As can be seen from Fig.~10~(a), the gap located at the bottom of the 
spectrum of the d--wave pair correlator cannot be identified
using SDM with  $L=20$ MVM only, because the averaged distance between 
two (SDM) $\delta$--peaks (vertical lines) is of the same order of magnitude.
Thus the true gap structure of the spectrum remains hidden. 
By choosing the MEM, one obtains a smooth reconstruction of the whole 
spectrum and achieves a significant better resolution of the gap 
from the same number of MVM. This becomes particularly obvious by  
looking at the $\omega$--integrated spectral function 
$N^{\mit \Delta}(\omega)$. In $N^{\mit \Delta}(\omega)$
the dominant low--energy QP peak 
(corresponding to the jump--like behaviour of $N^{\mit \Delta}(\omega)$
at the bottom of the spectrum) is clearly separated from 
the incoherent high--energy part by an energy gap 
(with  $N^{\mit \Delta}(\omega)\simeq const$). 
Of course, for both methods the energy resolution gets better  
with increasing number of moments 
(see Fig.~10~(b)), and consequently the SDM and MEM results 
converge on each other. However, since the dimension of the Hilbert space
($D=10^3$) is rather small for the spectra shown in Fig.~10~(b), 
we have calculated about 10~\% of the true peaks 
by using 120 MVM (on the other hand, for our EP systems 
having a total dimension of $10^9$ this corresponds to 
only $10^{-7}$ of the possible excitations).
Most notably, the spectral weight of different parts of the 
MEM spectrum does not change any more going from 20 to 120 MVM. 

Since for any finite system the true spectral function will 
consist of a sum of delta functions, we compare the efficiency of
KPM and MEM in dependence of the number of moments by reconstructing 
a fictitious reference spectrum of five $\delta$--peaks. 
Figure~11 demonstrates that the MEM provides a dramatic 
improvement in energy resolution over the KPM (typically by 
a factor of four). Thus the spectral
function is much better reconstructed, in particular 
when the number of MVM is small. On the other hand, an advantage of the  
linear KPM over non--linear MEM is that the 
KPM smoothing is uniquely determined by the number of 
Chebyshev recursions~\cite{SRVK96}, i.e., in contrast to the
MEM we have a uniform energy resolution $\propto \tilde{M}^{-1}$.

To analyze the accuracy of the KPM and MEM in more detail we have
compared in Fig.~12 a portion of a reconstructed single hole spectrum 
with the true spectral function obtained without moment truncation,
i.e., from the full diagonalization of the system.
Again convergence to the exact data is seen to be more rapid 
by applying the MEM (note that $\tilde{M}=4\hat{M}$). 
Moreover, the peaks in the KPM spectra are surrounded by 
not fully damped Gibbs oscillations. As can be seen from 
Fig.~12~(b), the MEM produces artificial peaks as well but
they are at least one order of magnitude smaller than those for KPM.
That means, these isolated spurious eigenvalues have extremely
small spectral weight and hence negligible influence on the 
cumulative spectrum. To illustrate that these spurious peaks 
can be easily identified, we have plotted the MEM spectrum 
for different number of moments. Contrary to the true
eigenvalues the spurious peaks (often labeled ``ghosts'') 
move about as $\hat{M}$ increases.
\narrowtext\def\baselinestretch{0.95}
\bibliography{ref}
\bibliographystyle{phys}

\figure{FIG. 1. Wave--vector resolved spectral functions of one hole 
in the t--J  (a) and Holstein t--J  (b) models. Results are obtained 
for a tilted $\sqrt{10}\times\sqrt{10}$ cluster with periodic boundary
conditions  ($\vec{K}$--vectors are given in units of $(\pi/5,\pi/5)$). 
The integrated densities of states are depicted 
in the lower panels of (a) and (b). The units of $A_{\vec{K}}$, 
though arbitrary, are nevertheless the same in all $\vec{K}$ sectors
displayed in  Figs.~1--3, so that the peak heights may be compared.}
\figure{FIG. 2. Single--hole spectral function $A_{\vec{K}}(\omega)$
and integrated spectral weight $N(\omega)$ for the 2D HtJM at $\omega_0=0.8$
for different values of $\ep$ and $J$. Note that the lowest peak 
in (c) and (d) contains contributions from each $\vec{K}$--sector.}
\figure{FIG. 3.  $A_{\vec{K}}(\omega)$ and $N(\omega)$ for the 2D HtJM  
in the non--adiabatic strong EP coupling regime ($\omega_0=3.0$, $\ep=4.0$).} 
\figure{FIG. 4. Band dispersion (a) and spectral weight factors (b) 
for the 2D HtJM on a 16--site lattice ($J=0.4$).}
\figure{FIG. 5. Spectral weight factor factor $Z_{\vec{K}}$ 
as a function of the EP coupling strength $\ep$ 
for the different $\vec{K}$ vectors of the ten--site lattice:  
(3,1) (triangles down), (2,4) (squares), (0,0) (circles), and
(5,5) (triangles up) [in units of ($\pi/5,\pi/5$)]. 
The inset shows the narrowing of the coherent 
bandwidth ${\mit \Delta}E$ with increasing $\ep$.}
\figure{FIG. 6. Optical conductivity in the 2D (Holstein) t--J model
for $J=0.4$. $\sigma^{reg}(\omega)$ and $S^{reg}(\omega)$ are obtained
for the ten--site lattice with 15 phonons. The single--hole ground state
has momentum $\vec{K}=(3\pi/5,\pi/5)$.}
\figure{FIG. 7. Optical conductivity in the 2D HtJM for $J=0.1$ and
$\omega_0=0.8$. Results are given at $\ep=1.0$ (a) and $\ep=4.0$
in the  $\vec{K}=(0,0)$ and $\vec{K}=(3\pi/5,\pi/5)$ sectors, respectively.}
\figure{FIG. 8. Convergence of the phonon distribution 
function $|c^m|^2(M)$ and ground--state
energy $E_0(M)$ (inset) as a function of the maximum number of 
phonons $M$ for the 2D HtJM on a ten--site lattice ($J=0.4$).
$E_0(M)$ is given with respect to the variational estimate of
ground--state energy, $E_0^{(IMVLF)}$, obtained 
within the IMVLF--Lanczos approach developed 
recently by the authors~\protect\cite{FRWM95}.}
\figure{FIG. 9. Dynamical d--wave pair correlation function
of the pure 2D t--J model, $A^{\mit \Delta}(\omega)=
\sum_n|\langle {\mit \Psi}^{(N-2)}_n|{\mit \Delta}|
{\mit \Psi}^{(N)}_0\rangle|^2\,\delta[\omega-(E^{(N-2)}_n-E^{(N)}_0)]$\,,
where ${\mit \Delta}=\frac{1}{\sqrt{N}}\sum_{i,\delta=\pm} 
\tilde{c}_{i\uparrow}(\tilde{c}_{i+\delta x\downarrow}-
\tilde{c}_{i+\delta y\downarrow})$. 
Results are calculated on a 16--site lattice with two holes for
$J=0.3$ using  SDM with $L=32$ and $64$ MVM.}  
\figure{FIG. 10. SDM (thin solid lines) and MEM (thin dashed lines) 
results obtained for the pair correlation function 
$A^{\mit \Delta}(\omega)$ in the 2D t--J model with 20 (a) and 120 
MVM (b) [$J=0.3$, $N=16$]. 
Shown is also the corresponding integrated spectral weight 
$N^{\mit \Delta}(\omega)=\int_{-\infty}^{\omega}d\omega^{\prime}
A^{\mit \Delta}(\omega^{\prime})$ (bold lines).}
\figure{FIG. 11. Comparison of KPM (a) and MEM (b) by means of a 
fictitious spectrum consisting of five $\delta$-peaks.
The reconstructed spectral function is shown
using $64$ (dashed),  $256$ (dotted), and $512$ (solid) moments.}
\figure{FIG. 12. Depicted is a part of the excitation
spectrum of a single hole with momentum $\vec{K}=(3,1)\pi/5$ 
in the 2D t--J model $(D=126)$. In upper panel, the KPM and
MEM reconstructed spectral function $A^{\tilde{c}}(\omega)$ 
has been calculated using $512$ and 128 moments, respectively.
For comparison the exact peak positions and spectral weights
are shown (stars). In the lower panel we illustrate how spurious 
peaks can be identified.}
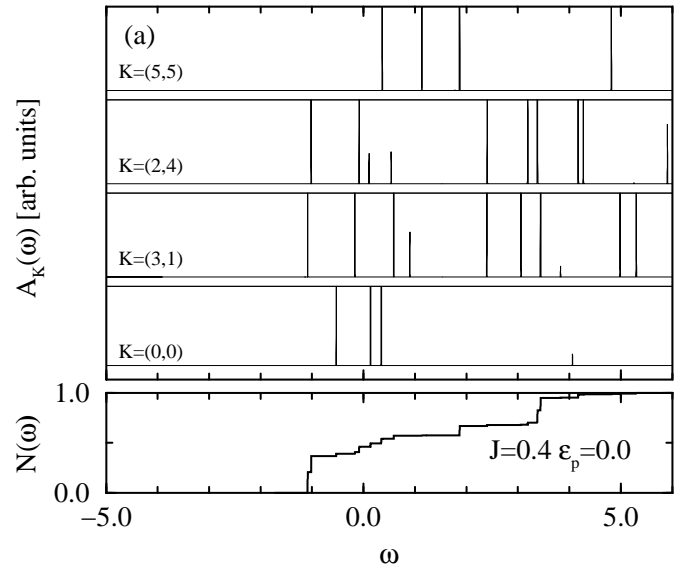
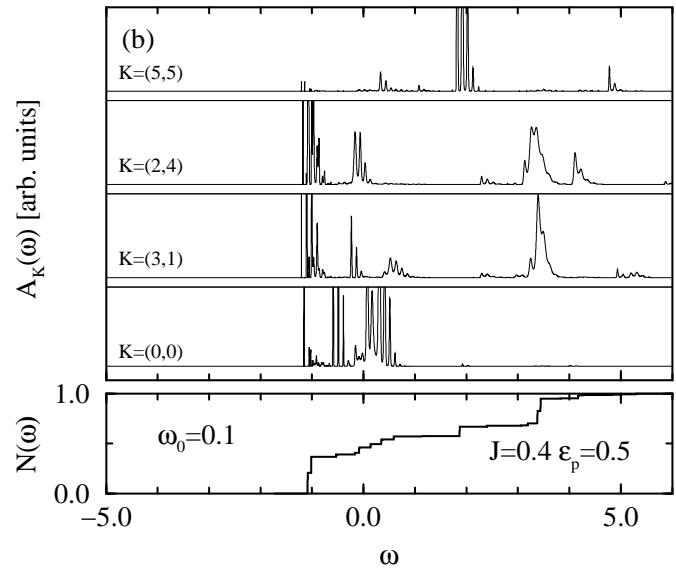
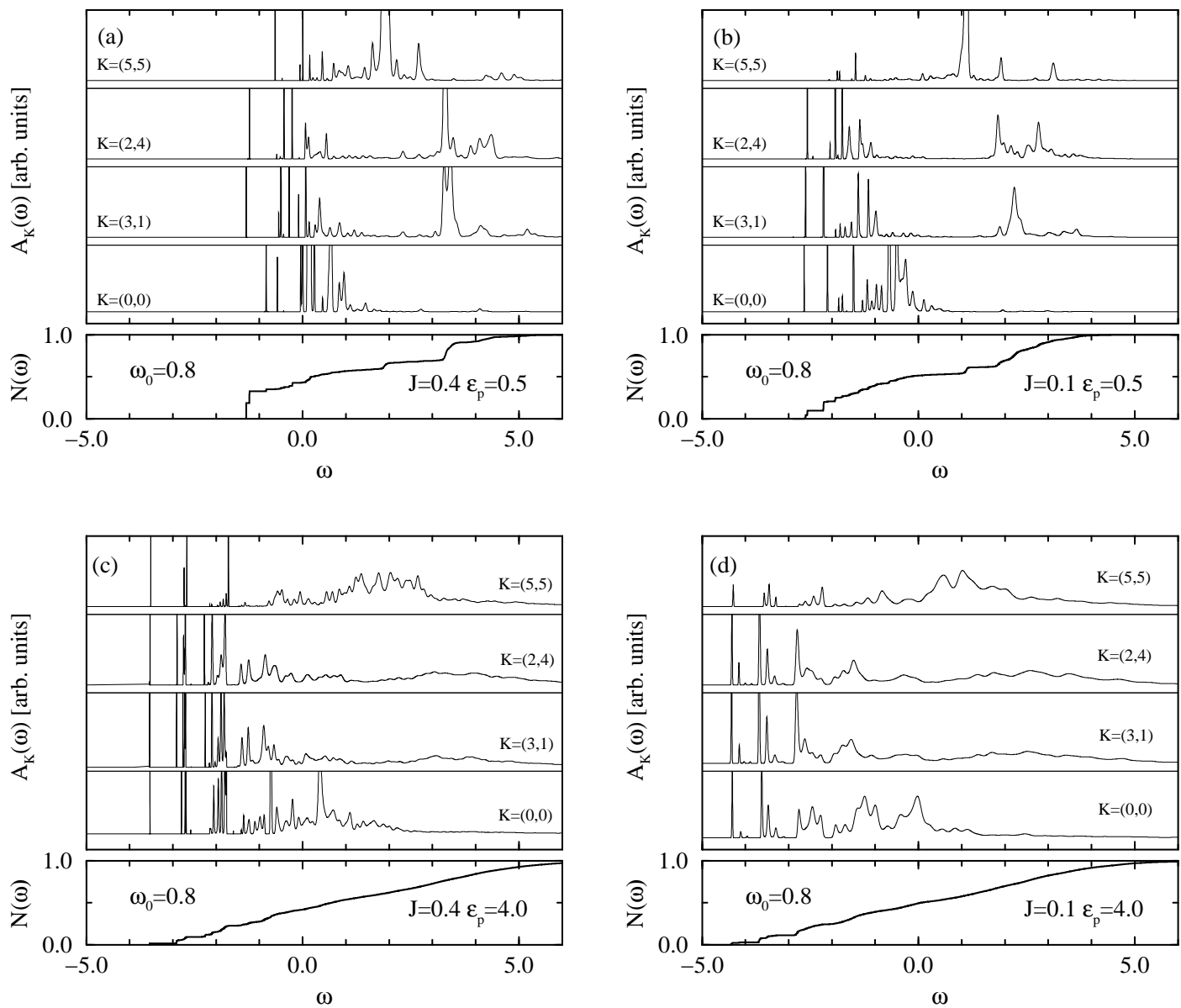
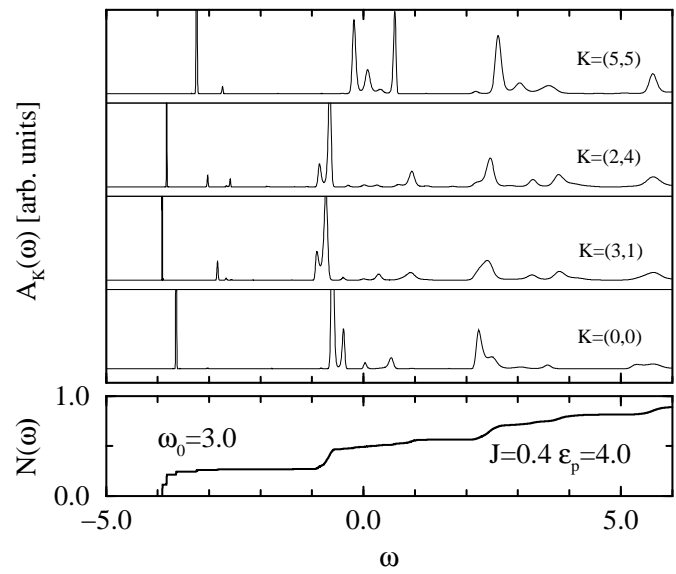
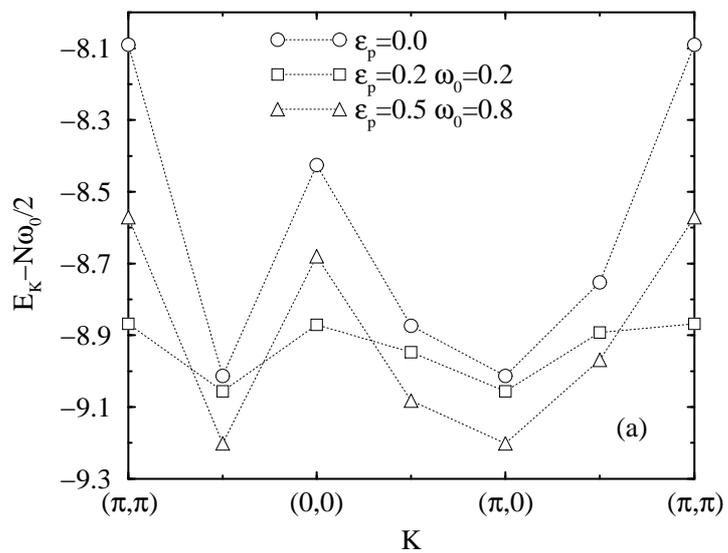
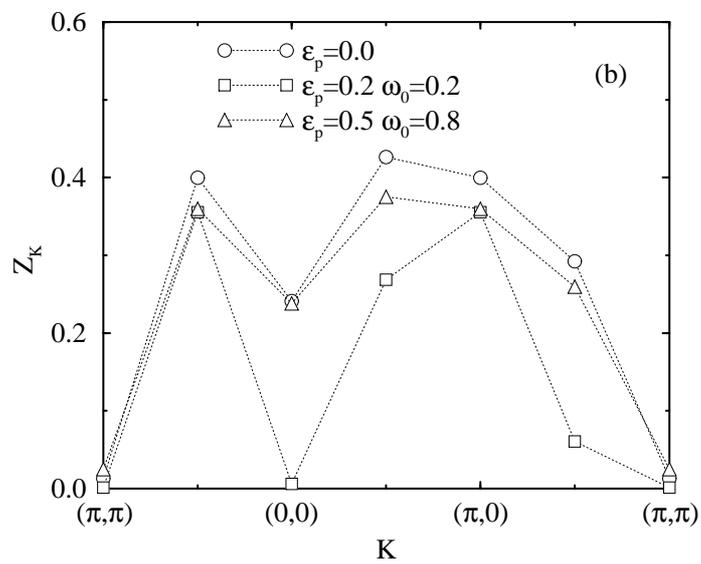
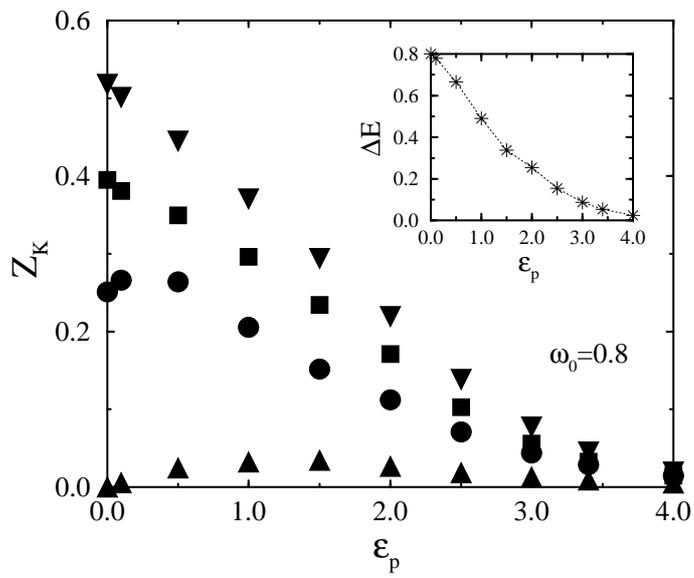
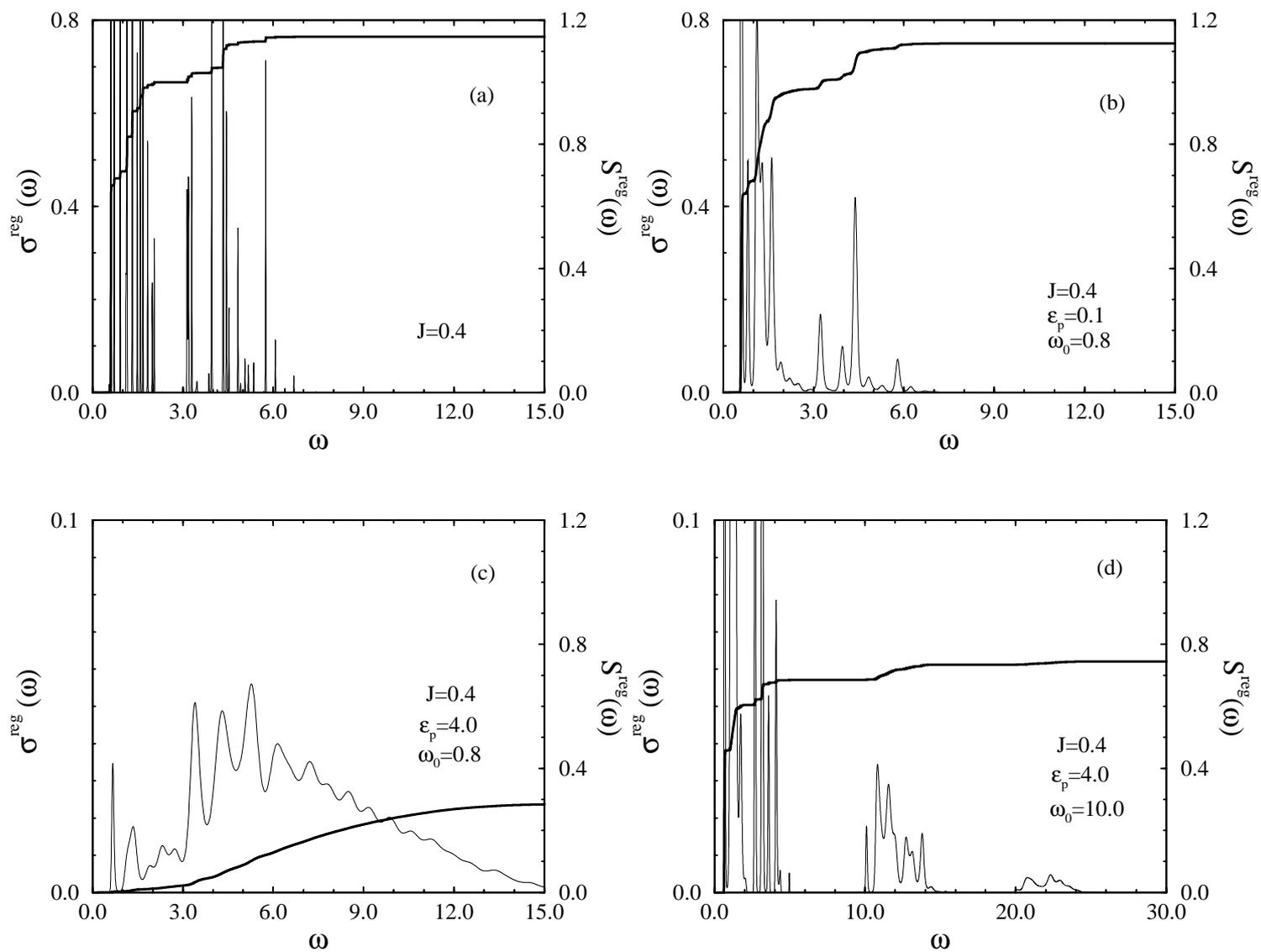
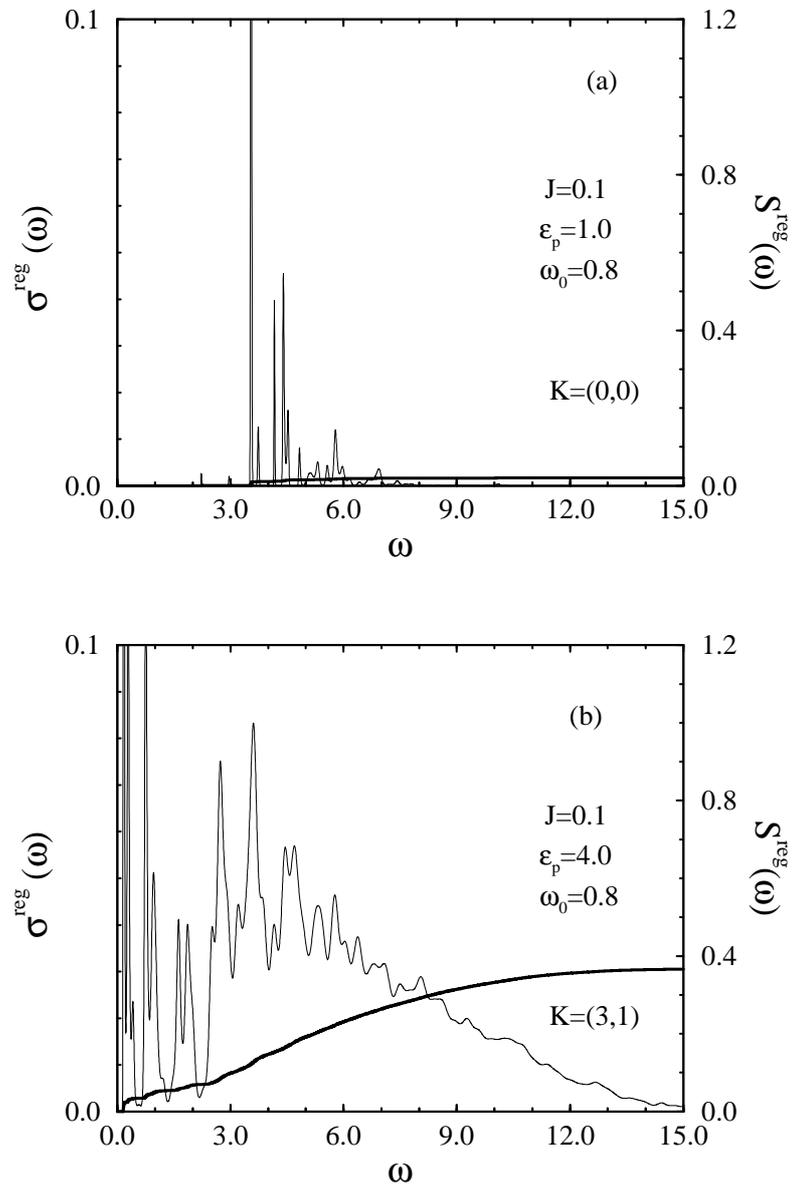
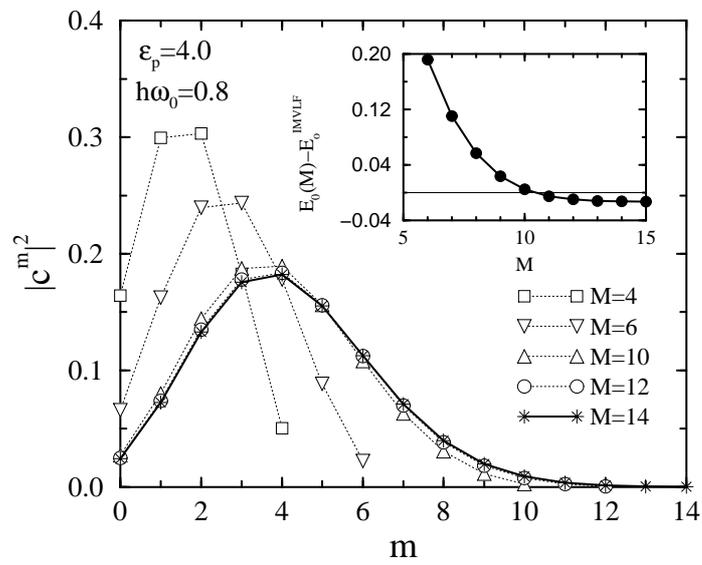
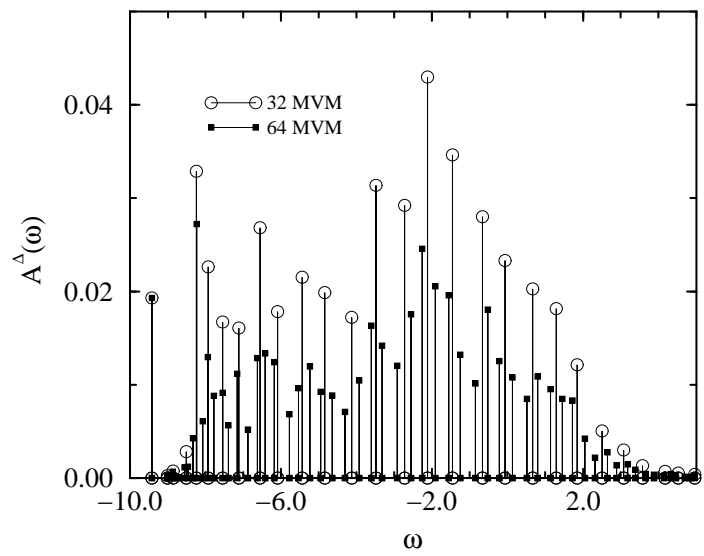
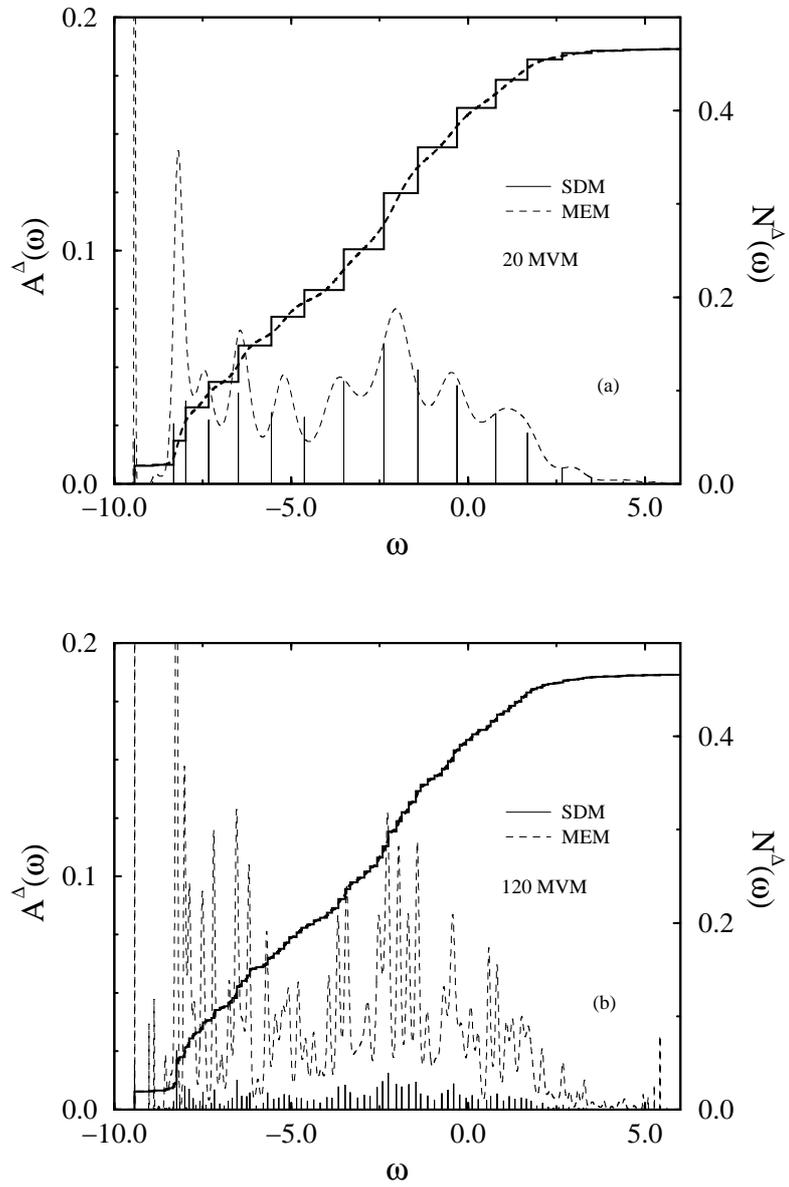
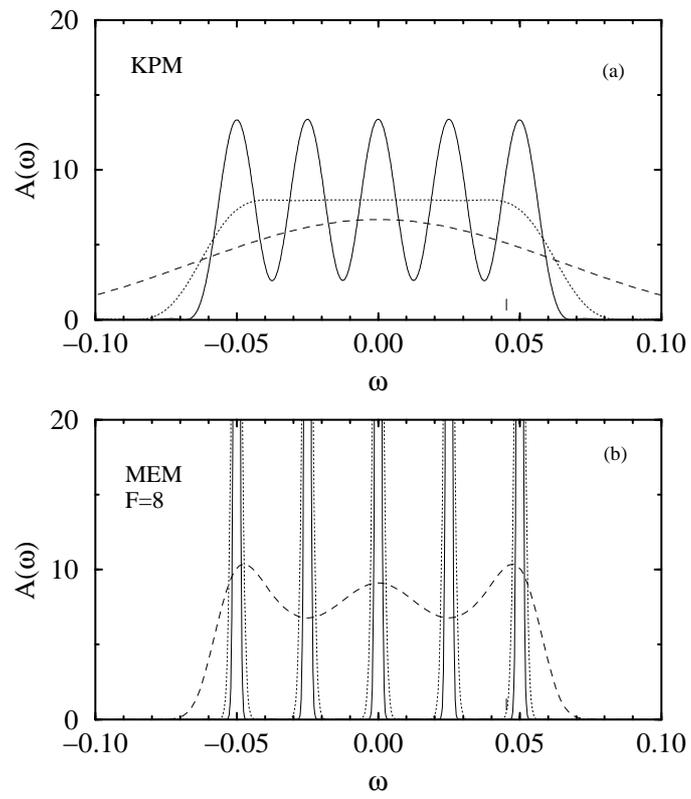
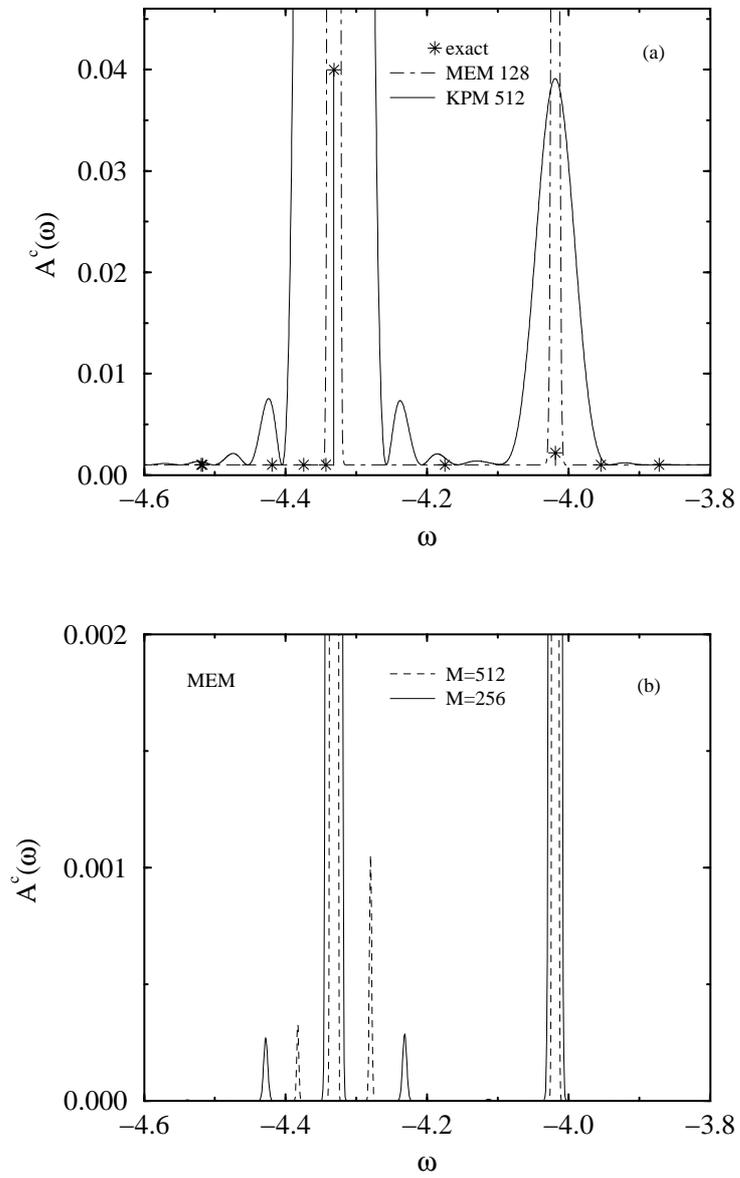
\end{document}